**Manuscript**

# Ecosystems in the Anthropocene: transformative drivers

Clara de Góes Monteiro de Carvalho Guimarães[1]

Pablo José Francisco Pena Rodrigues [1*]

[1]Instituto de Pesquisas Jardim Botânico do Rio de Janeiro, Rua Pacheco Leão, 915, CEP 22460-030, Rio de Janeiro, Brazil

*Corresponding author: Rodrigues, P.J.F.P. ( pablo@jbrj.gov.br )
Orcid: 0000-0001-8054-3643

**Abstract**

Human activity has an enormous impact on Earth, changing organisms, environments and landscapes, leading to the decline of original ecosystems and irreversible changes that create new combinations of living beings and materials. As a result, ecosystems with new properties and new species pools are emerging. Here, we explore a set of transformative drivers, which can act either individually or in synergy. The expansion of novel ecosystems — hybrids of natural and agricultural systems — is a sign of irreversible, human-induced change. Human growth, adaptation to climate change, urban expansion and geoengineering are powerful transformative drivers which are expected to have a high impact, creating novel ecosystems. In contrast, less transformative drivers such as degrowth, biocentrism, ecological restoration and low-impact agriculture can mitigate human impacts, leading to adaptation, resilience and sustainability, while conserving original ecosystems. This requires a new approach, incorporating new ecological, ethical and cultural perspectives, to keep ecosystems functional and healthy.

# 1 Introduction

The transformation caused by the Anthropocene is the world's greatest change triggered by a single species, causing major changes to ecosystems (Hobbs et al. 2006; Albuquerque et al. 2018; IPCC 2021). Humans have altered the Earth's extensively (Western 2001; Chure et al. 2022), and this trend is expected to continue beyond the fossil fuel era (Bardi 2016). Changes in nutrient cycles, species distributions and habitats (Vitousek et al. 1997; Tilman et al. 2001; Simberloff et al. 2013) may be irreversible, decreasing biodiversity (Vitousek et al. 1997; Folke et al. 2012) and resulting in the homogenization and simplification of communities (Parra-Sanchez et al. 2025). Human expansion has transcended boundaries (Niva et al. 2023), thereby facilitating species movement (Mack and Lonsdale 2001; Hobbs et al. 2006). This can result in new combinations of species and new properties emerging (Hobbs et al. 2006; Ellis and Ramankutty, 2008; Chure et al. 2022), making it often impossible to determine whether the change was intentional or random. This pattern began in the Neolithic (Ruddiman, 2007) and accelerated in the 20th century (Steffen et al. 2015).

The collapse of the Earth's systems by crossing planetary boundaries is possible (Scheffer et al. 2001; Steffen et al 2015), but it is also plausible that the planetary system will reach new equilibrium points (Rockström et al. 2009; Barnosky et al. 2012; Pinheiro and Pena-Rodrigues, 2025). There is a growing consensus that changes to ecosystems give rise to new combinations of organisms and materials (Hobbs et al. 2009; Gómez Márquez 2025). Their key characteristics include the potential to alter ecosystem functioning through new interactions (Osmolovsky et al. 2025). Ecosystems resulting from direct or indirect human action may not require continuous intervention and can be considered natural despite not being original (Illy and Vineis 2024). They can emerge in response to induced conditions and new factors (e.g., soil degradation, nutrient input, and alien introduction). This includes sites managed or induced, such as agroforestry and agricultural fields (Pretty 2008). New communities are now commonly found in managed ecosystems (Seastedt et al. 2008) and these changes have resulted in the local extinction of original species, followed by the introduction of new ones. Indeed, urban, cultivated, or degraded landscapes create dispersal barriers (Forman 1995), while creating novel ecosystems. Direct human impacts, such as soil removal, dam construction, harvesting and pollution, as well as indirect, such as erosion and overgrazing, generally lead to a decrease in biodiversity (Vitousek et al. 1997). These ecosystems represent an intermediate category between original and managed ones (Sanderson et al. 2002). The proportion of each type of ecosystem varies over time and space, with these variations often being driven by increases or decreases in human activities that are responsible for the anthropogenic flow of materials and energy. Other potential human actions include solar radiation management, which involves reflecting sunlight, and the injection of stratospheric aerosols to mimic the effect of volcanic eruptions by reflecting sunlight (Crutzen 2006). These techniques may have side effects, such as changes in precipitation and ozone layer damage (NRC 2015). Human activities are connected to the non-human flows of materials and energy that regulate the entire biosphere. Although our activities have great transformative power, they will always be subject to the limitations imposed by Earth's systems. Ultimately, we are just another species that depends on this balance to survive, and we must reconsider our transformative actions.

## 2 Humans, Conservation and Restoration

Current perspectives on the human way of life, as well as on conservation and restoration, are being re-evaluated (Archer et al. 2024), as preserving or restoring original ecosystems may no longer be viable (Hobbs et al. 2009; Wakefield 2018; Abhilash 2021; Scheffran et al. 2024). As both mitigating and reversing abiotic and biotic changes are difficult, the aim is often to keep a functional hybrid state (Simberloff et al. 2013) and increase the regenerative capacity of ecosystems (Illy and Vineis 2024). Novel ecosystems can be culturally valuable (Marris 2011) and systems with minimal changes can support hybrid systems (Choi 2004). Furthermore, the chance of human intervention (positive or negative) is increasing (Zhong et al. 2025). Changed ecosystems affect humans, economies, and livelihoods (Adger et al. 2005) and create opportunities, such as ecotourism, sea farming and sustainable fishing (Bulleri et al. 2018). But these changes can also negatively impact agriculture (Cinner et al. 2018), which threatens food security. Ocean acidification and global warming affect species availability (Gentry et al. 2021). Humans often need to adapt in response to changes in ecosystems to ensure socio-economic sustainability (Füssel and Klein 2006). Environmental changes can also lead to people's displacement and increased conflicts over resources and can amplify social and economic vulnerabilities (Adger et al. 2005). Collaborative approaches are essential to ensure the equitable management of socio-economic resources (Berkes et al. 2003). To address the challenges and opportunities of conservation and restoration in the transformative Anthropocene, an integrated approach involving research and active public participation is essential (see Fig. 1).

## 3 Novel ecosystems, species and materials

The distinction between novel and original ecosystems is not easy, and the criteria include human action as the trigger and self-perpetuation (Hobbs et al. 2013). This concept should be approached with caution, to ensure it is used as an opportunity for conservation and resource management and not a free pass for impacts (Miller and Bestelmeyer 2016). Spatially, it is crucial to determine if an ecosystem is novel (Perring et al. 2014), and over time, the historical baseline can vary depending on location and culture. Historical baselines are easier to determine in pristine sites than in places with long use. Agriculture, urbanization and climate change can lead to a transition, with the arrival of new species and materials altering ecosystems. This often has a negative impact on communities, causing changes in biomass, biogeochemical cycles and resource availability (Table 1).

Ecosystems are changing, and the future is uncertain, particularly considering the interactions between species and materials. Organisms favored by humans, such as transgenics, hybrids and aliens, will increasingly become part of ecosystems and will introduce new properties (Pena-Rodrigues and Lira 2019). If a new community maintains or enhances functional diversity, it may help to preserve ecosystems. Furthermore, new types of human-generated materials can influence interactions, create new living environments and alter biological properties. Such as in artificial reefs, which harbor high levels of biodiversity (Vivier et al. 2021), and in managed terrestrial ecosystems (e.g. cities, agroforests and greenhouses).

## 4 Nature-based solutions and Carbon sequestration

Combining carbon sequestration with Nature-based solutions is an efficient way of mitigating the environmental and climate crises. Tree planting and algae cultivation are ways of sequestering carbon. Other methods consist of capturing and storing it in reservoirs (Riahi et al. 2004). Nature-based solutions offer ecological benefits, including enhancing biodiversity, improving water quality, reducing the impact of disasters, and sequestering carbon (Seddon et al. 2020). While the potential is high, ecological mismatches and unintended impacts on biodiversity are common. Success depends on research, thoughtful implementation, and governance (Lavigne de Lemos et al. 2024). Carbon credits offset CO2 emissions (Riahi et al. 2004: Albert et al. 2021). However, it depends on the integrity of the actions and the ability to report and verify the reductions (Figueroa et al. 2007; Dubey and Arora 2022). Genuine and effective measures must be taken to address long-term sequestration (Table 1).

## 5 Agricultural transitions

Agricultural transition is urgent yet complex; intensifying and diversifying agriculture can change ecosystems (Galiana et al. 2017). Agroforestry, biological control and sustainable management offer a way forward towards less harmful agriculture (Figure 1). Overproduction of food causes population growth and ecological problems (Chatti and Majeed 2025). Animal farming is a source of greenhouse gases, contributing to global warming, acidification and eutrophication (Djekic 2015). Plant-based diets have a lower environmental impact (Baroni et al. 2007). Indeed, sustainable agriculture balances food production and ecosystems (Tilman et al. 2001) and the transition to sustainable practices is driven by technology (Foley et al. 2005). Precision agriculture can reduce the environmental impact by optimizing the use of water and fertilizer (Padhiary et al. 2024), while biological control replaces synthetic pesticides (Van Lenteren et al. 2012; Ayilara et al. 2023). Crop rotation, conservation agriculture and integrated nutrient management are low-impact actions (Pretty 2018). Sustainable intensification can preserve ecosystems by decreasing agricultural areas (Foley et al. 2005).

## 6 Degrowth and Biocentrism

Environmental ethics lies between the anthropocentric perspective, which treats nature as a resource, and non-anthropocentric view (Nash 1989). By contrast, the non-anthropocentric biocentrism and degrowth advocate the intrinsic value of nature (Devall and Sessions 1985) emphasizing the moral consideration of nature as a subject (Naess 1989). Modern civilization led to human growth, environmental degradation, social division and crises (Danish et al. 2019). Degrowth is rooted in a critique of the perpetual economic growth, which is environmentally unsustainable, exacerbates inequalities, and compromises ecosystem health (Kallis et al. 2018). Human choices have ecological consequences, especially for vulnerable populations (Martinez-Alier 2002). We need to change the way we produce and consume to ensure that resources are shared more fairly. Economic growth has natural limits determined by the carrying capacity of ecosystems (Daly1996). Degrowth also means rethinking society to make it inclusive (Salleh 2009) and sustainability can be achieved through equitable decision-making processes (Schlosberg 2007). The integration of perspectives points to a future in which human well-being and ecosystem health are inseparable and interdependent. Degrowth and biocentrism proposes a radical change to the basis of our social and economic systems.

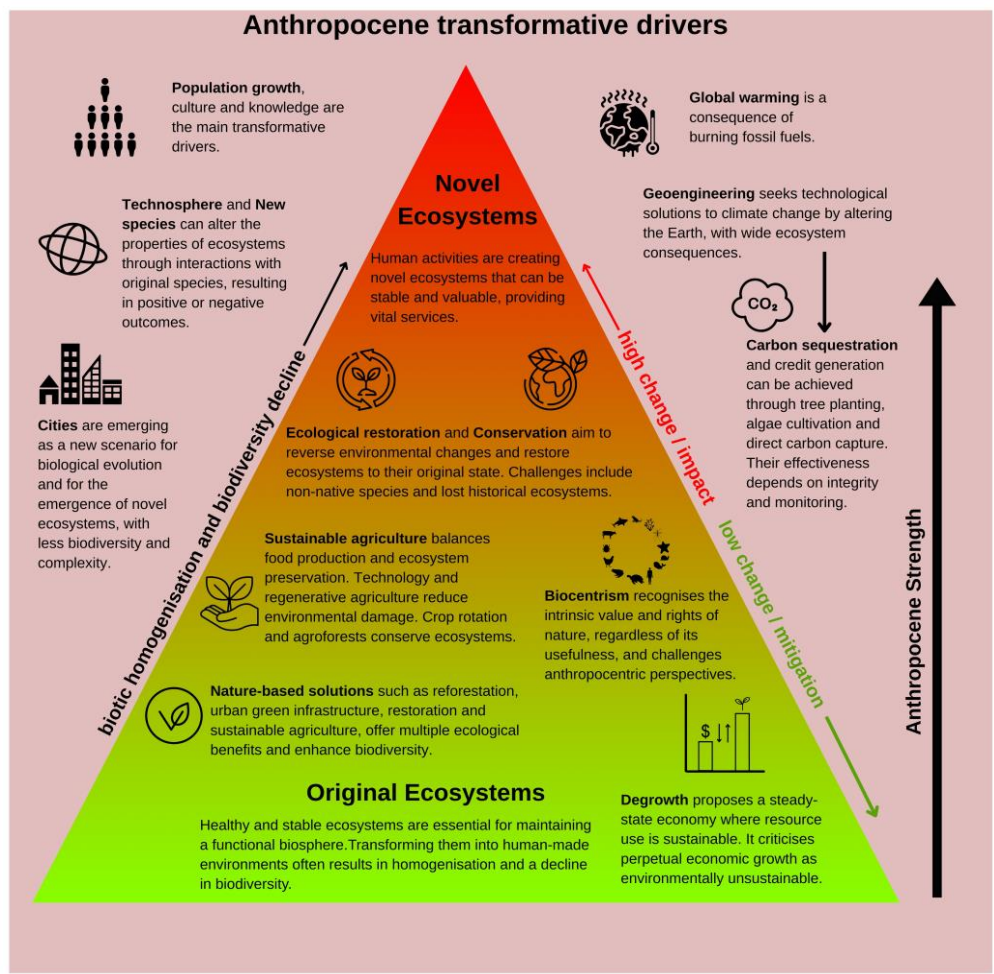

**Fig.1** Selection of a set of drivers that can lead to the simplification and homogenization of ecosystems or reduce human impact. As the strength of the Anthropocene increases, changes

caused by new flows of energy and materials are expected to transform ecosystems. Drivers with a high potential for conserving original ecosystems are at the base of the pyramid, while those with high transformative potential are at the top and around it. In extreme scenarios, anthropogenic changes will result in the formation of novel ecosystems. This figure also illustrates the potential for a reduction in the number of ecosystems, which could be replaced by cities, agricultural areas and other human-made ecosystems.

**Table 1** Selection of eight Anthropocene transformative drivers, along with their potential perspectives and key aspects.

| Transformative drivers | Perspectives | Key aspects | References |
|---|---|---|---|
| Cities | Building resilient cities is essential, but it has the potential to transform ecosystems | Adapt, transform and persist to maintain functions | Meerow and Newell 2016; Liu et al. 2025 |
| | | Integrating disaster risk reduction into development practices | UNDRR 2015 |
| | | Planned urbanization can mitigate impacts | Shamsuddin 2020 |
| | | Urban resilience needs land use planning and public housing | Tortajada 2014 |
| | | Broad development agendas ensure cities adapt to change | World Bank 2013 |
| Novel ecosystems | Adaptive strategies for managing both transformed and original ecosystems that are facing human intervention | Human activity and self-perpetuation as critical triggers | Hobbs et al. 2006; Hobbs et al. 2013 |
| | | Historical baselines are impossible in sites with a long use | Wakefield 2018 |
| | | Human activities and climate change: transition to new | Hobbs et al 2011 |
| | | Novel ecosystems are temporary evolutionary stages | Perring et al. 2014 |
| | | The interplay between novel, original and managed systems | Marris 2011 |
| | | They can influence the provision of ecosystem services | Mace et al. 2012 |
| | | Change species distribution and abundance | Hobbs and Huenneke 1992 |
| | | Human actions are changing ecosystems | Ellis et al. 2021; Edgeworth et al. 2025 |
| Species and materials | New communities preserve ecosystems and human-generated materials will influence interactions | New species and materials can alter ecosystem properties | Vitousek et al. 1996; Simberloff et al. 2013 |
| | | New species combinations: changes in biomass and resources | Mack and Lonsdale 2001 |
| | | Species can introduce new properties to an ecosystem | Tilman 1997 |
| | | Functional redundancy – similar roles in sites near agriculture | Tilman et al. 2012 |
| | | Functional diversity can be crucial for ecosystem resilience | Hooper et al. 2005 |
| | | Technosphere can create new habitats and change habitats | Zalasiewicz et al. 2016 |
| | | Human-made structures can harbor high levels of biodiversity | Vivier et al. 2021 |
| Nature-based solutions | Enhance biodiversity, reduce human impact and sequester carbon | Reforestation, restoration, and sustainable agriculture | Cohen-Shacham et al. 2016 |
| | | Science, thoughtful implementation, and governance | Lavigne de Lemos et al. 2024 |
| | | Climate change mitigation to keep global warming below 2°C | Griscom et al. 2017; Seddon et al. 2020 |
| Carbon sequestration | Genuine and effective actions to avoid adverse outcomes | Carbon certificates that represent a reduction in $CO_2$ | Albert et al. 2021 |
| | | Carbon capture: storing it in reservoirs | Riahi et al. 2004 |
| | | Carbon credits only work if the actions are verified | Figueroa et al. 2007; Dubey and Arora 2022 |
| | | Permanence (long-term sequestration) is a key concern | Anderson et al. 2019 |
| Agricultural transitions | New technologies and sustainable management can protect ecosystems | Sustainable agriculture balances production and ecosystems | Tilman et al. 2001 |
| | | Sustainable practices driven by technology keep ecosystems | Foley et al. 2005 |
| | | Precision agriculture can reduce environmental impact | Padhiary et al. 2024 |
| | | Biological pest control can replace synthetic pesticides | vanLenteren et al. 2012; Ayilara et al. 2023 |

**Conclusions**

In the face of immense uncertainty, improving our understanding of the anthropogenic transformation of ecosystems is crucial. Human activities have resulted in the emergence of novel ecosystems that exhibit unique characteristics and dynamics. Managing novel and original ecosystems requires an understanding of their complexities and the integration of new ecological, socio-economic, ethical, and cultural perspectives. These ecosystems, along with agricultural fields and cities, must be incorporated into new conservation and adaptation strategies. Adaptation and mitigation that have high transformative potential, such as geoengineering, carbon credit and resilient cities, need to be planned to avoid deleterious effects. Perspectives integrated into the natural functioning of ecosystems — such as nature-based solutions, agroforestry, and biocentrism — can help reduce the impact of our activities, making them less detrimental and transformative.

Author Contributions: Both authors contributed equally to the preparation of this paper and have read and agreed to the published version of the manuscript.

Funding: This research received no external funding

Informed Consent Statement: Not applicable.

Data Availability Statement: Not applicable.

Conflicts of Interest: The authors declare no conflict of interest.